\documentstyle[12pt,aaspp4,epsfig]{article}

\begin{document}

\title{Breaking the Disk/Halo Degeneracy with Gravitational Lensing}

\author{Ariyeh H. Maller\altaffilmark{1},
Luc Simard\altaffilmark{2},
Puragra Guhathakurta\altaffilmark{2,3},
Jens Hjorth\altaffilmark{4},
Andreas O. Jaunsen\altaffilmark{5},
Ricardo A. Flores\altaffilmark{6},
and Joel R. Primack\altaffilmark{1}}
\altaffiltext{1}{Physics Department, University of California, Santa Cruz,
 CA 95064, USA}
\altaffiltext{2}{UCO/Lick Observatory, Astronomy and Astrophysics 
Department, University of California, Santa Cruz, CA 95064, USA}
\altaffiltext{3}{Alfred P. Sloan Research Fellow}
\altaffiltext{4}{Astronomical Observatory, University of Copenhagen,
Juliane Maries Vej 30, DK-2100 Copenhagen {\O}, Denmark}
\altaffiltext{5}{Institute for Theoretical Astrophysics, Oslo, Norway}
\altaffiltext{6}{Physics Department, University of Missouri, Saint Louis,
 MO 63121, USA}
\begin{abstract}
The degeneracy between the disk and the dark matter contribution
to galaxy rotation curves remains an important uncertainty in our understanding
of disk galaxies.
Here we discuss a new method for breaking this degeneracy using gravitational
lensing by spiral galaxies,
and apply this method to the spiral lens B1600+434 as an example.
The combined image and lens photometry constraints allow models for B1600+434
with either a nearly singular dark matter halo, or a halo with a sizable core.
If the dark halo has a core,
then the bulge dominates the gravitational potential in the inner part 
of the galaxy, its mass is between 1.3 
and $1.5 \times 10^{11} M_{\sun}$,
and the disk mass is less then $5 \times 10^{10} M_{\sun}$.  
If the dark halo is singular,
there is a degeneracy between the disk mass 
and the halo ellipticity. The dark halo flattening ($c/a$) can be as low as  
0.53 if there is no disk mass, while the maximum allowed disk mass, 
$1.3 \times 10^{11}M_{\sun}$, is reached with a spherical halo. 
{ \it A maximum disk model is ruled out with high confidence.}
Further information, such as the circular velocity of this galaxy, will help
break the degeneracies. Future studies of spiral galaxy lenses will be able to
determine the relative contribution of disk, bulge, and halo to the
mass in the inner parts of galaxies.   
\end{abstract}


\section{Introduction}
The discovery that rotation curves do not fall off near their optical edge
\markcite{rtf:78}({Rubin}, {Thonnard}, \& {Ford} 1978) nor beyond \markcite{bosma:78}(Bosma 1978) remains the strongest 
evidence for the existence of dark matter on galactic scales.
Still, how much dark matter 
contributes to the gravitational field in the inner parts of galaxies 
remains an open question 
\markcite{bosma:98,sell:98}(see {Bosma} 1998; {Sellwood} 1998, for recent reviews).
One complication, and 
an unresolved issue in itself, is the three dimensional shape of dark
matter halos.  Dark halos are usually 
assumed to be spherical in fitting rotations curves, 
yet found to be ellipsoidal in N-body simulations \markcite{dc:91,warr:92}({Dubinski} \& {Carlberg} 1991; {Warren} {et~al.} 1992)
and in a variety of observational studies 
\markcite{sack:98}(for a recent review see {Sackett} 1998).
In this paper we will demonstrate that constraints on the mass of the disk and 
the ellipticity of the dark halo can be found by studying spiral galaxies that
are strong lenses.

The uncertainty in the relative contributions of the 
disk, bulge, and halo to the circular velocity is due to our lack 
of knowledge about the mass-to-light ratios of the disk and bulge, 
and the profile of the dark halo.  One tactic has been
to assume the contribution from the disk to the rotation curve 
is as large as possible; this ``maximum'' disk hypothesis 
\markcite{vs:86}({Van Albada} \& {Sancisi} 1986) then favors a dark halo with a large core radius. Whether 
or not this is a good assumption is a topic of much 
debate \markcite{cr:99,sp:99}({Courteau} \& {Rix} 1999; {Sallucci} \& {Persic} 1999). 
Unfortunately, we can not determine the mass-to-light ratios of the 
baryons from first principles because they are 
extremely sensitive to assumptions 
about the low mass end of the initial mass function (IMF) \markcite{worthey:94}({Worthey} 1994).
If we knew the dark matter profile we could determine the mass-to-light ratios
by fitting rotation curves and hence constrain the low mass end of the IMF.
Thus, identifying the baryonic contribution to the inner rotation curve places
important constraints on the IMF, and the profile of dark matter halos.
Galaxy formation models \markcite{dss:97,mmw:98,sp:98}(e.g. {Dalcanton}, {Spergel}, \&  {Summers} 1997; {Mo}, {Mao}, \& {White} 1998; Somerville \& Primack 1999) assume an IMF
and a dark halo profile to predict a range of disk masses for a given 
halo mass.  Thus any measurement of the baryonic mass fraction in a spiral 
galaxy will test these models.

We will demonstrate how gravitational lensing of spiral galaxies
can be used to help determine the relative contributions of the 
disk, bulge, and halo to
the inner parts of spiral galaxy rotation curves.  The basic premise is that 
gravitational lensing is sensitive to the total projected mass and to the 
combined ellipticity of the mass distribution.  Since the projected 
ellipticities of disk, bulge, and dark halo are in general different, it is 
possible to constrain their combination. 

In Section \ref{disks} we 
summarize some aspects of lensing by disks.  In Section \ref{mass} we 
discuss the mass distribution in spiral galaxies and how to measure it with 
gravitational lensing.  In Section \ref{b16} we introduce B1600+434, an 
edge-on spiral galaxy, and discuss its optical properties.  
In Section \ref{lens} we study B1600+434 as a gravitational lens
to see what constraints can be placed on its mass distribution.
We end with some conclusions and a discussion of the future of this
method.

\section{Lensing by Disks} \label{disks}

\begin{figure}[tp] % figdens
\vskip .5pc
\centering
\centerline{\epsfig{file=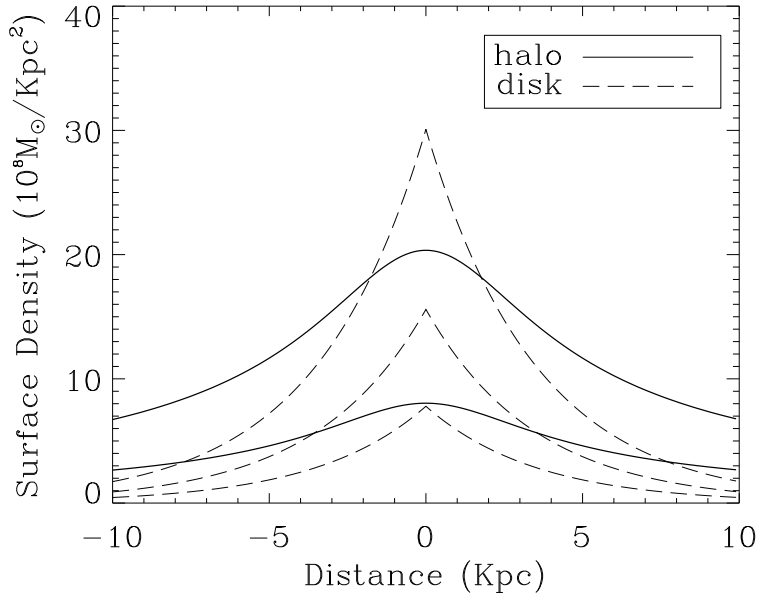,height=7cm}}
\vspace{10pt}
\caption{
Shown is the projected density of an isothermal sphere (solid) 
with a core radius equal to the
disk scale length and circular velocities of 220 and 280 km s$^{-1}$ 
(bottom to top), and the projected density of a exponential disk (dashed) 
at an inclination of 
$0\arcdeg,60\arcdeg$ and $80\arcdeg$ (bottom up).  The disk properties 
are chosen to match the Milky Way.
}\label{figdens}
\end{figure}

Gravitational lensing by the disks of spiral galaxies has only recently
received attention.
Originally, spiral galaxies were thought 
to contribute only 10-20\% 
of the total lensing cross section \markcite{tog:84}({Turner}, {Ostriker}, \& {Gott} 1984).
\markcite{loeb:96}{Loeb} (1996) and \markcite{mfp:97}{Maller}, {Flores}, \& {Primack} (1997) noticed that the 
projected surface mass density of a nearly edge-on spiral galaxy
is comparable to that of an elliptical galaxy 
with a higher velocity dispersion (see Figure \ref{figdens}).
This leads to an increase of the area inside the inclination averaged caustic
of roughly 50\% \markcite{mfp:97,wt:97}({Maller} {et~al.} 1997; {Wang} \& {Turner} 1997) and a corresponding increase
in the expected number of spiral lenses.  Spiral galaxies have a
higher cross section for high magnification \markcite{mb:98}({M\"{o}ller} \& {Blain} 1998) which, 
when the magnification bias is taken into account, can lead to an increase
in the optical depth as large as a factor of three depending on how one 
models the evolution of disk properties \markcite{bmm:99}({Blain}, {M\"{o}ller}, \& {Maller} 1999).
However, most of the studied lenses have been identified as early type galaxies
\markcite{kkf:98}({Keeton}, {Kochanek}, \& {Falco} 1998), and do not display the ``disk''
morphology in which both images are located to one side of the lens 
\markcite{kk:98}({Keeton} \& {Kochanek} 1998). A reasonable explanation is that spiral lenses 
become fainter due to obscuration by dust and thus are not detected in
optical surveys \markcite{bl:98}({Bartelmann} \& {Loeb} 1998).
Strong evidence for this hypothesis exists based on the fact that five out
of six identified spiral lenses were discovered in the CLASS radio survey
\markcite{browne:98}({Browne} 1998), for which dust obscuration is unimportant.

Two other important effects have not been considered in determining 
the total optical depth in spiral galaxy lensing.  
First, the luminosity function, the starting point for determining 
optical depth, is not corrected for extinction. The Tully-Fisher relation 
\markcite{tf:77}({Tully} \& {Fisher} 1977), which relates luminosity to circular velocity, includes an 
inclination correction, and thus a distribution of inclinations must be 
included when converting luminosities to circular velocities.
When this is done correctly, one finds that massive spiral galaxies inhabit
halos with velocity dispersions comparable to those of
ellipticals \markcite{gonzales:99}(Gonzales {et~al.} 1999).  In addition, galaxies are not isolated 
systems, and many gravitational lenses are members of groups or small clusters.
This should also be taken into effect when calculating the total optical depth.

Regardless of their total optical depth, spiral lenses offer new
opportunities to study the distribution of mass in spiral galaxies 
\markcite{mfp:97,kk:98}({Maller} {et~al.} 1997; {Keeton} \& {Kochanek} 1998).
\markcite{mfp:98}{Maller}, {Flores}, \& {Primack} (1998) demonstrated that spiral lenses could be 
used to study the fraction of mass in baryons and the shape and profile of 
the dark matter halo. \markcite{kdj:98}{Koopmans}, {De Bruyn}, \&  {Jackson} (1998) did the first 
detailed analysis of a spiral lens, the system B1600+434. 
They showed with numerical models that this spiral galaxy could
not have a dark halo with an axis ratio $c/a$ flatter then 0.5, but did not 
have adequate photometry to constrain the disk or bulge masses.  
We return to B1600+434 in Section \ref{b16}.

\section{Mass in Spiral Galaxies} \label{mass}
\subsection{Mass Components}

Spiral galaxies are composed of three mass components: disk, bulge, 
and dark halo (for our purposes we will not consider the contribution 
to the mass from a gaseous disk). 
We assume that spiral galaxies have azimuthal symmetry, and thus 
it is natural to use cylindrical coordinates with the disk lying in the 
$z=0$ plane.  Furthermore, we assume the mass distributions of the bulge
and halo have homoeoidal symmetry (the flattening $q_3=c/a$ 
remains constant) such that they can be expressed 
as functions of the elliptical radius $m_3$, where $m_3^2=R^2+(z/q_3)^2$.  
We will use the subscript 3 to distinguish a variable
in three dimensional space from quantities projected along the line of sight
into a two dimensional plane.  Two dimensional quantities are related to their
three dimensional counterparts by 
the projected ellipticity $q$, which is given by 
\begin{equation}
q^2=q_3^2{\rm{sin}}^2(i)+{\rm{cos}}^2(i).
\end{equation}
Here $i$ is the disk inclination and $i=0\arcdeg$ corresponds to a face-on 
disk.  In the projected plane of the sky 
the mass components become ellipses that can be 
represented as functions of $m$, where $m^2=x^2+(y/q)^2$.
Disks are well fit by an exponential profile, while bulges in projection
have been fit with a de Vaucouleurs profile \markcite{deV:48}(de~Vaucoulers 1948)
or an exponential profile \markcite{dejong:94,cdb:96,andredakis:98}({De Jong} 1994; {Courteau}, {De Jong}, \&  {Broeils} 1996; {Andredakis} 1998).
By fitting elliptical isophotes to an image of the galaxy we can 
measure the disk inclination $i$, the
projected bulge ellipticity $q_{b}$, and the disk and bulge scale lengths.

The amplitude of the dark halo
contribution to the circular velocity (outside the region where the disk and
bulge contributions are important) can be measured from the rotation curve,
but its core radius, $R_{h}$, and 
ellipticity, $q_{3h}$, are not directly observable.  Thus, there are nine 
parameters which determine the mass distribution in a spiral galaxy: a length, 
a mass, and an axis ratio (inclination in the case of the disk) 
for each component. Five can be 
directly measured, but four must be determined from other constraints.  
The inner part of the rotation curve effectively gives 
only one additional constraint as the dark halo profile is already chosen 
to produce the correct shape.  The rotation curve gives
no constraint on the dark halo flattening and thus leaves
a degeneracy between the disk mass, bulge
mass and dark halo core radius.  To break this degeneracy
one must measure the potential gradient
in an orthogonal direction providing a different 
combination of these parameters. One way to do this is with gravitational 
lensing.  

\begin{figure}[tp] % figellip
\vskip .0pc
\begin{minipage}[b]{.5\linewidth}
\centering
\centerline{\epsfig{file=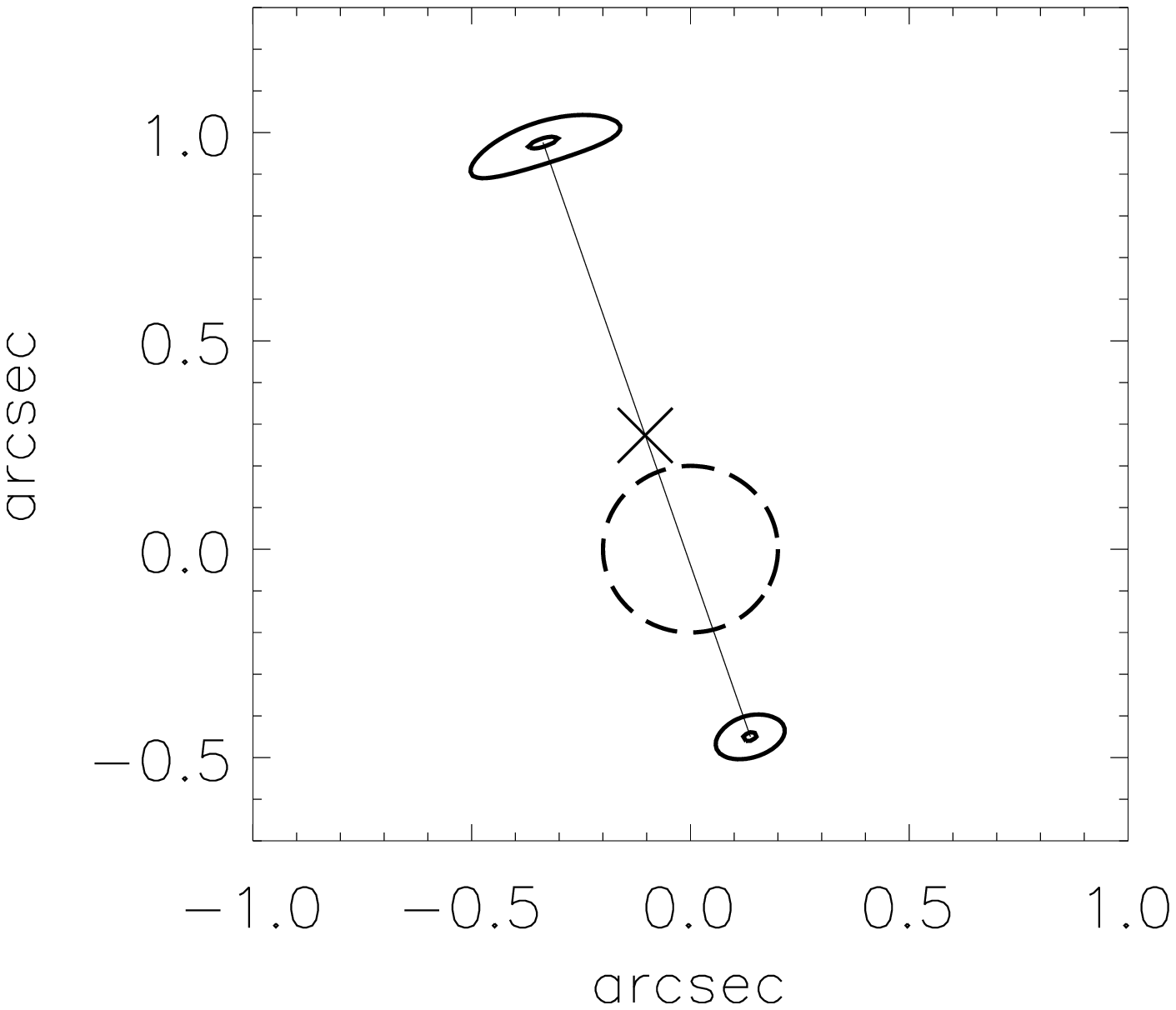,width=\linewidth}}
\end{minipage}\hfill
\begin{minipage}[b]{.5\linewidth}
\centering
\centerline{\epsfig{file=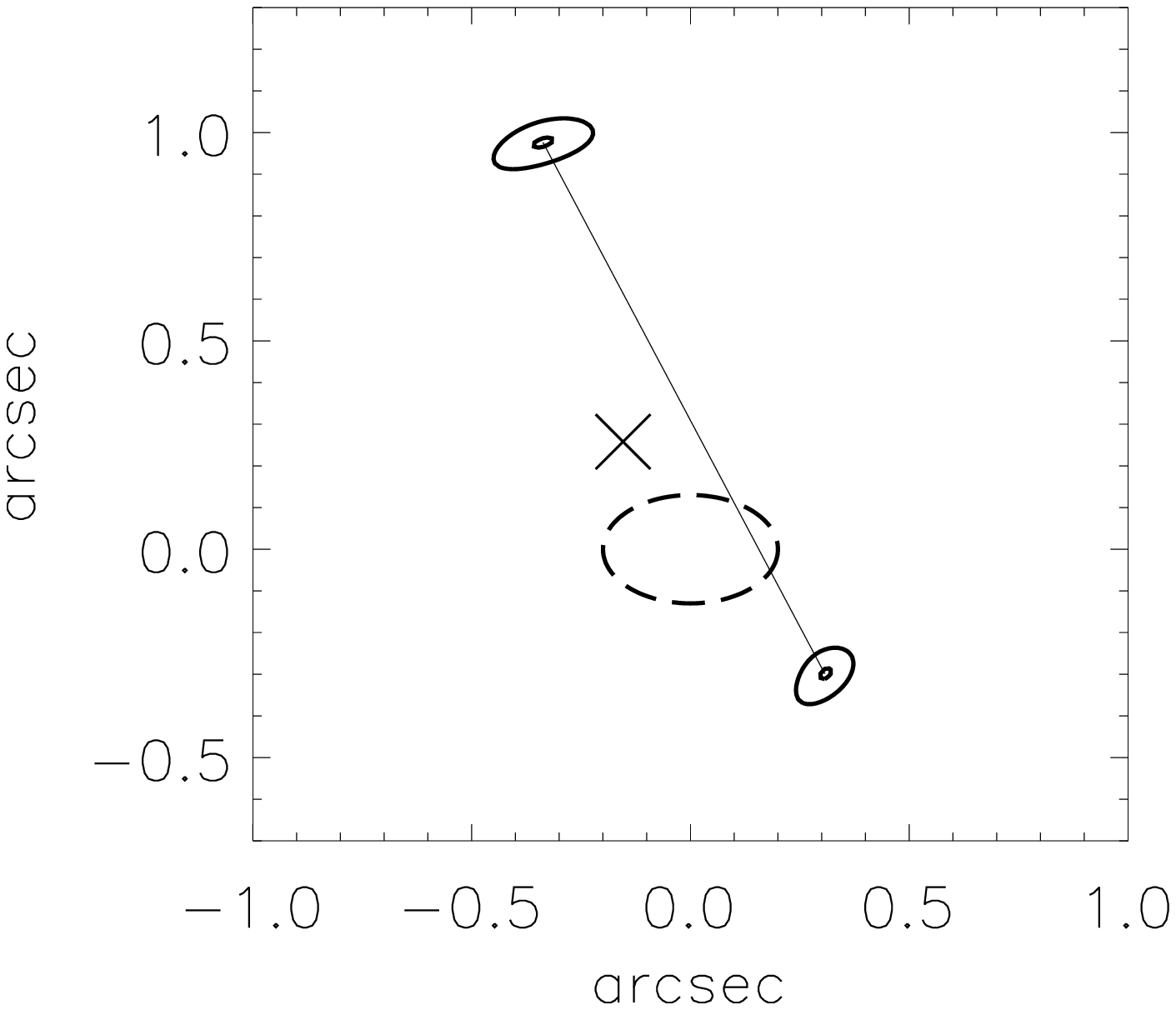,width=\linewidth}}
\end{minipage}\hfill
\vspace{10pt}
\caption{
Here we show the image configuration from a mass distribution that projects 
into a circle (left panel) and into an ellipse with axis ratio $q_h=0.65$ 
(right panel). 
The upper left image is held at a fixed position while the other image 
position is calculated.  The dashed line represents an isodensity contour of 
the lensing mass distribution.  
For the circular case the images can be connected 
by a line that passes through the lens center, but when the mass 
distribution is flattened the image position is displaced  to the right.
The X marks the source position.
}\label{figellip}
\end{figure}

\subsection{Mass from Lensing}

Gravitational lensing has become one of the primary tools for measuring
mass in astronomy \markcite{nb:96}({Narayan} \& {Bartelmann} 1999).  For galaxies,
lensing has been used to study the extent of dark halos
\markcite{hudson:98,dt:96,bbs:96}({Hudson} {et~al.} 1998; {Dell'Antonio} \& {Tyson} 1996; {Brainerd}, {Blandford}, \&  {Smail} 1996),
to measure the evolution of the fundamental plane
of ellipticals \markcite{koch:98}(Kochanek {et~al.} 1998), and the mass-to-light ratio of early type 
galaxies in clusters \markcite{nata:98}({Natarajan} {et~al.} 1998).  
In these cases the mass distribution is close to circular and 
the only property that can be rigorously determined is the total mass inside 
some radius.  With circular symmetry only the total mass enclosed determines
the gradient of the potential as stipulated by Gauss' Law.
In the case of edge-on spiral lenses the situation is different.

A circularly symmetric lens must produce images that are collinear with the 
center of the mass concentration.  
Ellipticity in the mass distribution pushes 
the line joining the two images to one side of the mass distribution, 
and the larger the ellipticity
the farther this line moves from the lens center.  
This is illustrated in Figure 
\ref{figellip} where we show the image configuration for an isothermal 
mass distribution with a spherical and with a flattened shape.
The position of one image is kept constant; with the introduction of 
ellipticity into the mass distribution, the other image moves away from the
lens center. 
If the ellipticity causes the source position to cross a caustic then 
four images will be produced whose positions can also be used to measure the
ellipticity of the mass distribution.
Thus with spiral lenses there are two strong constraints from
a lensing analysis: the total mass and the combined 
ellipticity of the projected mass.  
As we expect the disk, bulge, and halo to have different ellipticities, 
this gives a different constraint on their combination than the total 
mass alone. Since we can not measure the ellipticity of the dark halo, there 
exists a degeneracy between a more massive disk and a more flattened halo;
however, as we will show in the case of B1600+434, these can be separated since
they contribute differently to the circular rotation at the outskirts of the
disk.

\section{B1600+434: Observations and Analysis} \label{b16}

B1600+434 was first identified as a doubly imaged quasar with a source 
redshift $z_s =1.6$ and an image separation of 1\farcs4
\markcite{jack:95}({Jackson} {et~al.} 1995).  \markcite{jh:97}{Jaunsen} \& {Hjorth} (1997) identified the 
lens as an edge-on spiral galaxy. \markcite{fc:98}{Fassnacht} \& {Cohen} (1998) measured the 
lens redshift $z_l$ to be 0.41.  \markcite{kdj:98}{Koopmans} {et~al.} (1998) 
using an $I$ band WFPC2 image of the lens galaxy were able to show 
that the dark matter halo was no flatter then $q_{h} \geq 0.5$.  They were
not able to constrain the mass in the disk because the disk and bulge 
parameters were not extractable from the image photometry.

As part of the CASTLES project \markcite{munoz:99}(Munoz {et~al.} 1999), 
four 640 second images were taken with NICMOS's
NIC2 camera in the $H$ Band. These images were drizzled \markcite{fh:98}({Fruchter} \& {Hook} 1998) to give 
37.5 mas per pixel resolution. The images were fit with a PSF-convolved
2D bulge+disk model \markcite{simard:98,ms:98}(GIM2D: {Simard} 1998; {Marleau} \& {Simard} 1998)
to determine the optical properties of the galaxy, many to a few percent
accuracy (Table \ref{tableprop}).
We find that the bulge is well fit by an exponential profile, but not by 
a de Vaucouleurs profile.

\begin{figure}[tp] % figb16
\vskip .5pc
\begin{minipage}[b]{.46\linewidth}
\centering
\centerline{\epsfig{file=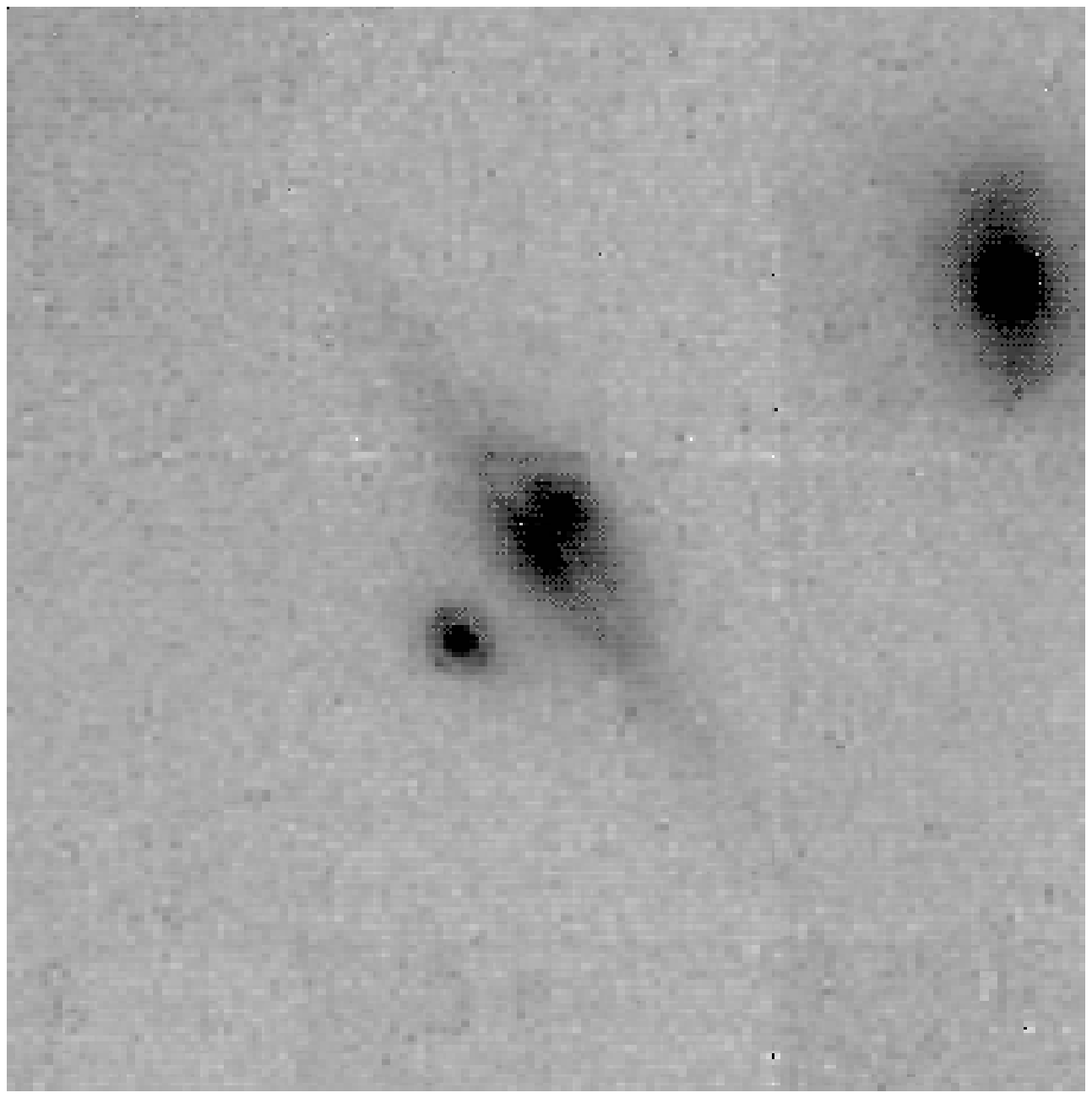,height=.9\linewidth,width=\linewidth}}
\end{minipage}\hfill
\begin{minipage}[b]{.46\linewidth}
\centering
\centerline{\epsfig{file=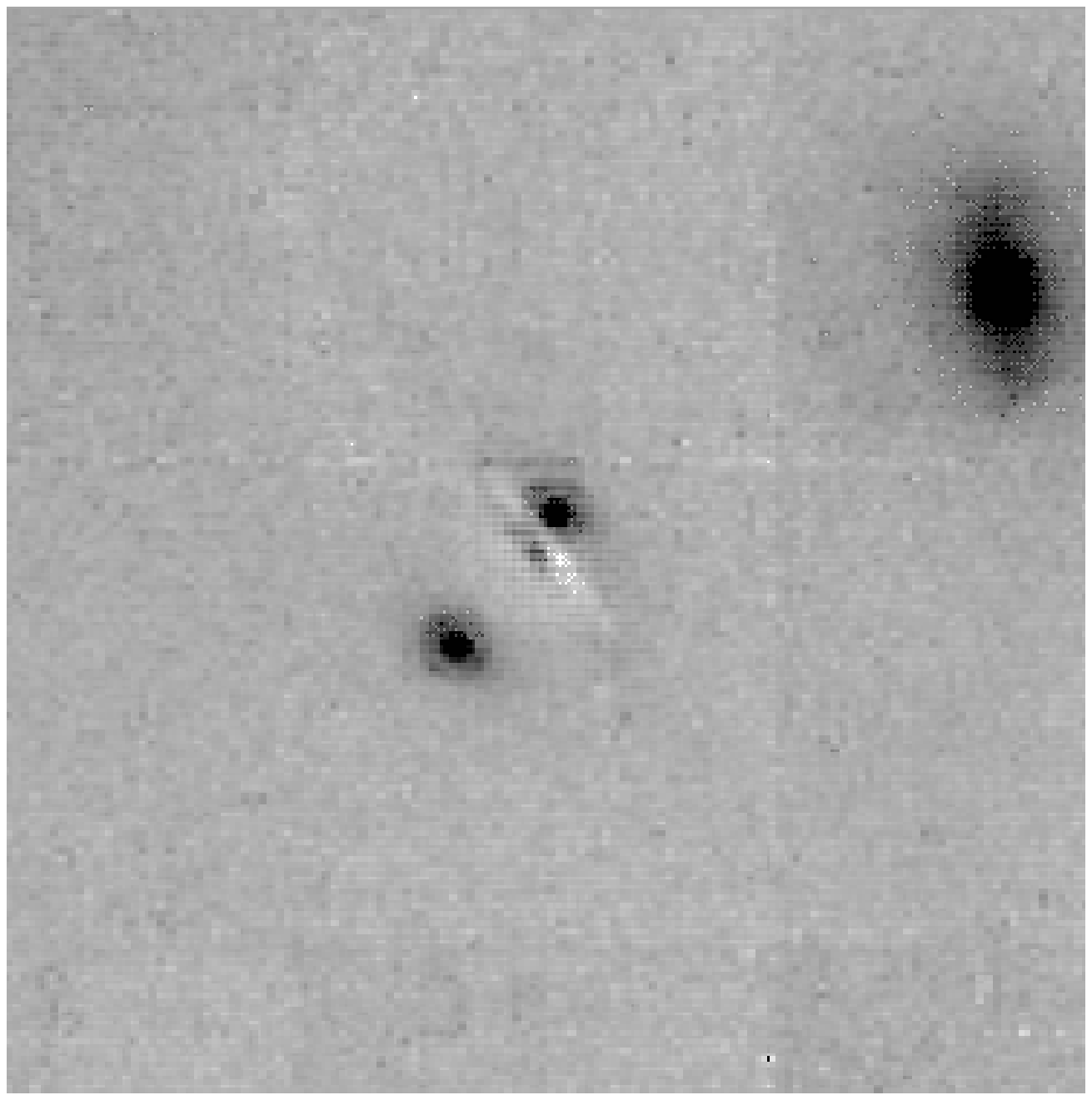,height=.9\linewidth,width=\linewidth}}
\end{minipage}\hfill
\vspace{10pt}
\caption{
The left panel shows the dithered NICMOS image of B1600+434
while the right panel shows
the same image with the lens galaxy surface brightness model
removed. The residual dust lane is clearly
seen. The companion galaxy is seen in the upper right corner. The images are 
9\arcsec on a side and taken as part of the CASTLES project. 
The NICMOS camera was rotated 186.07 degrees from North. So up and right
are roughly South and East.
}\label{figb16}
\end{figure}

Figure \ref{figb16} shows the NICMOS image and the same image 
with the galaxy subtracted.  
The pixels in the dust lane are masked out when using GIM2D to model the 
light distribution, so that when the lens galaxy surface brightness model
is subtracted from the image the dust lane produces a clear residual.  
Thus the results shown in Table 
\ref{tableprop} are with the dust lane removed. One also sees a companion 
galaxy 4\farcs35 from the center of the lens galaxy. 
This galaxy is well fit by a 
de Vaucouleurs bulge and an exponential disk and has an $H$-magnitude of 
17.3.

The total luminosity of 
the lens galaxy is 17.8 magnitudes in the observed $H$ band. To compare
with local galaxies we would like the rest frame luminosity of the 
lens galaxy in the I band. From \markcite{pogg:97}{Poggianti} (1997) we see that the K correction
and evolution correction for an Sa galaxy is -.2 and -.4 
magnitudes respectively, and the I-H color is 1.35 mag.
Taking $\Omega=1$ and H$_0=60$ km s$^{-1}$Mpc$^{-1}$ this implies an absolute 
magnitude $M_I=-23.18$ or 
$8.0 \times 10^{10} h_{60}^{-2} L_{\sun}$ where we have made 
no extinction correction.  We can then estimate the
rotation velocity of this galaxy using the $I$ band Tully-Fisher relation
\markcite{bern:94}({Bernstein} {et~al.} 1994).
Assuming no dust correction is necessary once  
the dust lane is removed, 
the corresponding Tully-Fisher velocity is $V_{TF}= 212$ km s$^{-1}$. If 
instead we make the full dust correction suggested by \markcite{bern:94}{Bernstein} {et~al.} (1994)
which amounts to 1.2 mag for this galaxy's inclination,
the Tully-Fisher velocity becomes $V_{TF}= 340$ km s$^{-1}$.
Either of these assumptions is probably extreme so we expect these values to 
bracket the true value. 
% We can also estimate the Tully-Fisher velocity for 
% the companion galaxy by the same method.  Assuming its redshift is the 
% same as B1600+434 the implied circular velocity is 265 km s$^{-1}$
% 251.4.

Table \ref{tableprop} shows that the disk and bulge PA (clockwise from 
North)  are not
perfectly aligned.  In this analysis we force the bulge and disk PA to be 
aligned and take the best value to be $34.4\arcdeg \pm 10\arcdeg$. 
Table \ref{tableprop} gives us four of the mass distribution parameters; 
unfortunately at this time there is no measure of the circular velocity of 
the galaxy, so there remain five unknown parameters.
We only get three strong constraints from the lensing analysis, so we expect 
a plane of solutions.

\begin{table}[b]     
\center
\caption{Optical Properties of the lens galaxy, B1600+434}
\label{tableprop}
\begin{tabular}{lccc}
\tableline
Property & best fit & 99\% lower bound & 99\% upper bound \\
\tableline
\tableline
Total Luminosity (mJy)   & 0.0990 & 0.0984  & 0.0999 \\
Bulge Fraction   & 0.5712  & 0.556  & 0.586  \\
Bulge Half Light Radius (arcsec)  & 0.504 & 0.494  & 0.514 \\
Bulge Ellipticity ($q_{b}$)   & 0.728 & 0.704  & 0.750 \\
Bulge PA (degrees)  & 24.4 & 27.1  & 22.9 \\
Disk Scale Length (arcsec)  & 1.35 & 1.33  & 1.36 \\
Disk Inclination ($i$) & 81.7 & 81.2 & 82.1 \\
Disk PA (degrees)  & 44.4 & 44.9  & 44.1 \\
\tableline
\end{tabular}
\end{table}

\section{Modeling B1600+434 as a gravitational lens} \label{lens}

\begin{figure}[t] % figcham
\vskip .5pc
\begin{minipage}[b]{.46\linewidth}
\centering
\centerline{\epsfig{file=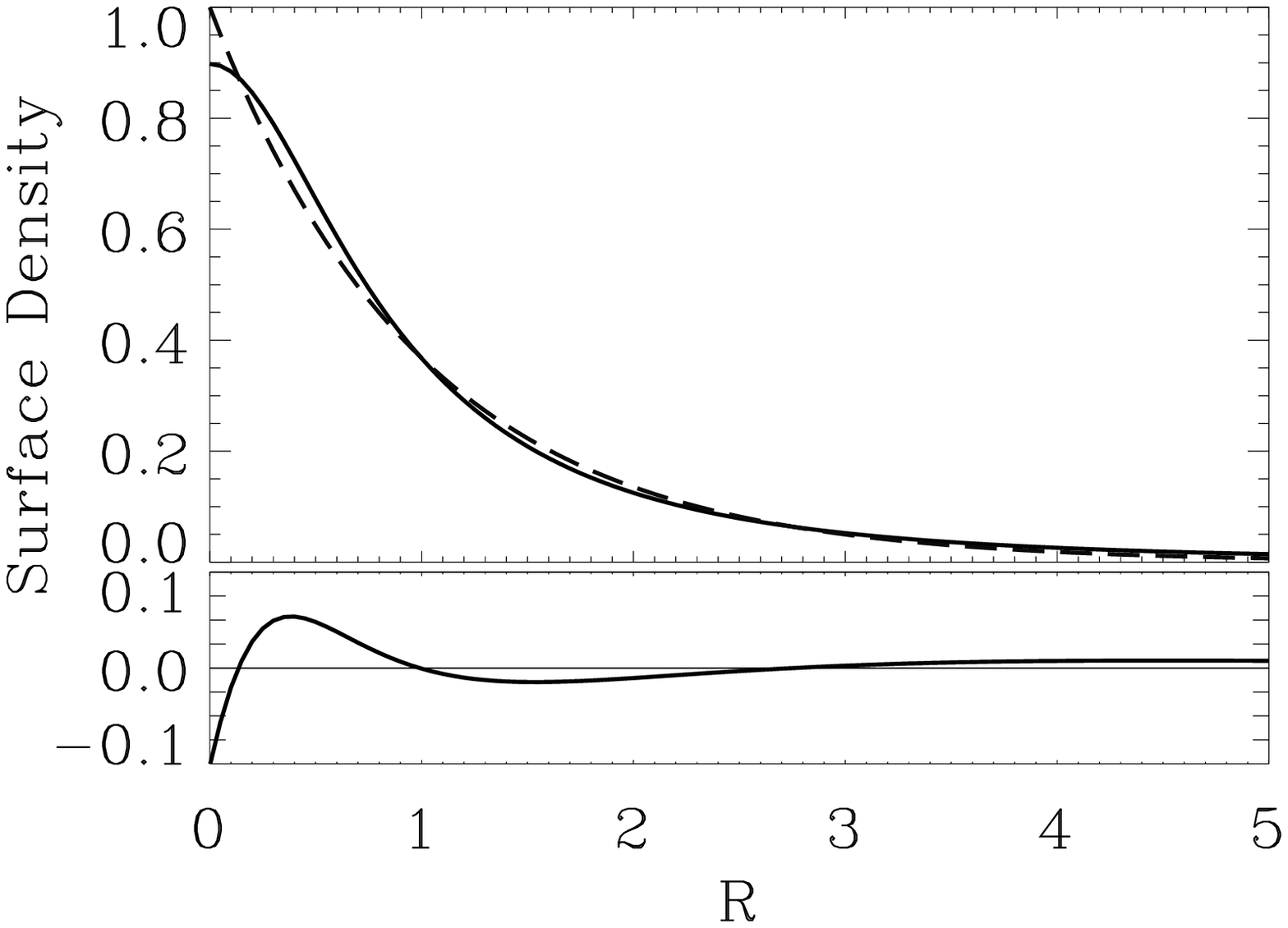,width=\linewidth}}
\end{minipage}\hfill
\begin{minipage}[b]{.46\linewidth}
\centering
\centerline{\epsfig{file=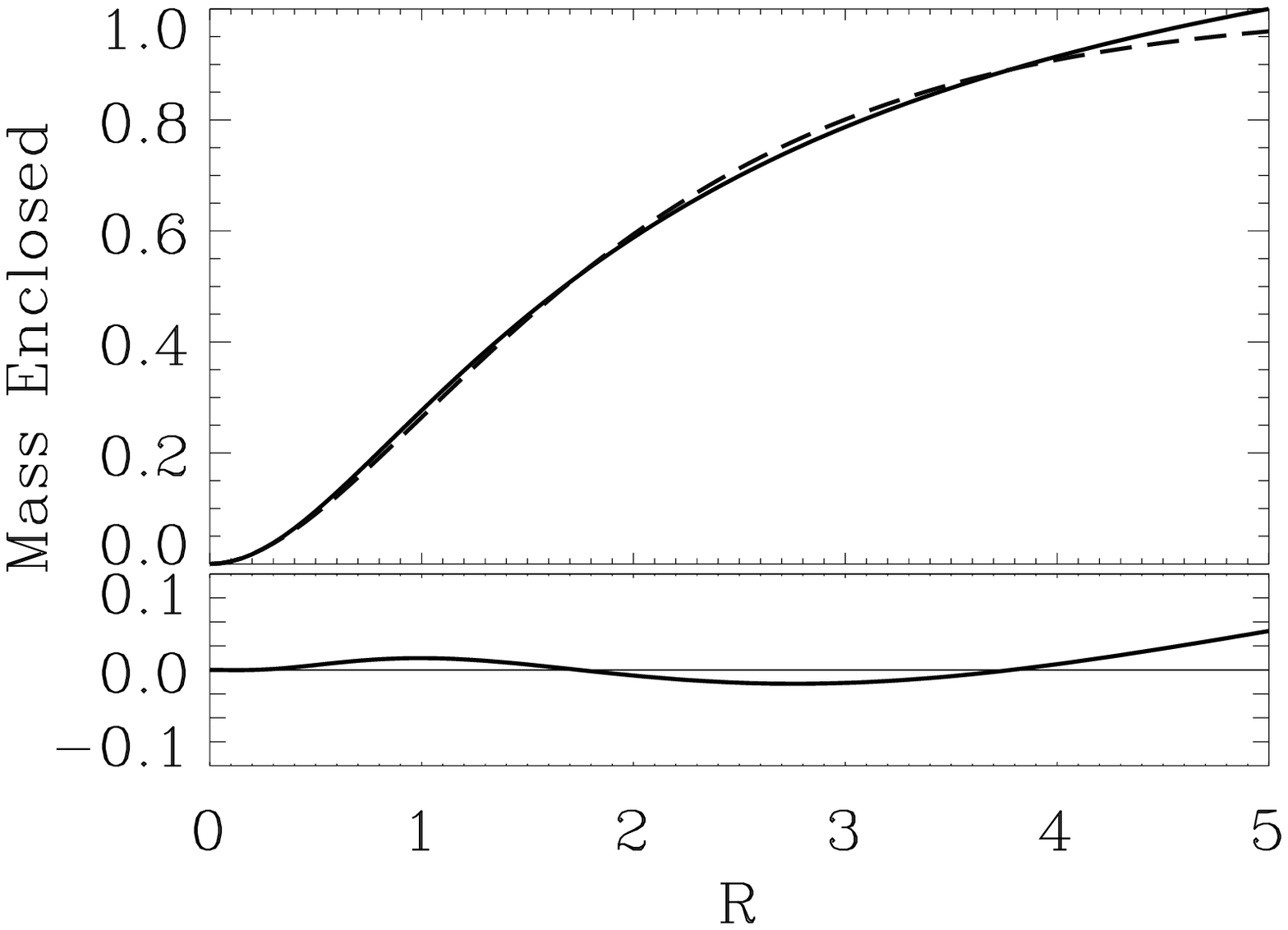,width=\linewidth}}
\end{minipage}\hfill
\vspace{10pt}
\caption{
An exponential profile (dashed) and a tuned 
chameleon profile (solid) are plotted in density (left) 
and projected mass (right) as a function of R,
the number of disk scale lengths.
The lower panel shows the difference between the two profiles. The
parameters of the chameleon are tuned to match the exponential by taking 
a=${\frac{1}{3}}R_d$ and s=${\frac{7}{3}}R_d$. 
}\label{figcham}
\end{figure}

There are five unknown parameters
in the mass distribution of this spiral galaxy: the 
masses of the disk and bulge,
or equivalently, their baryonic mass-to-light ratios,
the dark halo flattening $q_{3h}$, core radius $R_h$, and circular 
velocity $V_h$. To model the dark matter we will use the standard Pseudo 
Isothermal Elliptical Mass Distribution \markcite{kk:93,ksb:94}(PIEMD: {Kassiola} \& {Kovner} 1993; {Kormann}, {Schneider}, \&  {Bartelmann} 1994),
which has the form 
\begin{equation}
\rho(m_3)=\rho_0\frac{1}{{m_3^{2}+R_h^{2}}}.
\end{equation}
We use what we call the chameleon profile to model the exponential
profiles of the disk and bulge. This profile contains two scale lengths that 
can be tuned to mimic a number of commonly used density profiles
\markcite{maller:99,hk:99,kk:98}(Maller 1999; Hjorth \& Kneib 1999; {Keeton} \& {Kochanek} 1998).  Its surface density distribution is given by
\begin{equation}
\Sigma (m)=\Sigma _{0} \frac{as}{s-a} 
 \left( \frac{1}{\sqrt{m^{2}+a^{2}}}-\frac{1}{\sqrt{m^{2}+s^{2}}}\right)
\end{equation}
where choosing $a=\frac{1}{3}R_d$ and   
$s=\frac{7}{3}R_d$ closely follows an exponential
out to five scale lengths (Figure \ref{figcham}). Its lensing 
properties are analytic, and it deprojects into the three dimensional 
distribution 
\begin{equation}
\rho(m_3)=\frac{q}{\pi}\Sigma_{0} \frac{as}{s-a} 
 \left( \frac{1}{{m_3^{2}+a^{2}}}-\frac{1}{{m_3^{2}+s^{2}}}\right)
\end{equation}
which is just the difference between two PIEMDs.

\subsubsection{Confidence Levels}
A double image lens gives five constraints on a lensing model: four positions
and one flux ratio.  
As stated above there are five unknown parameters in the
mass model of the lens galaxy as well as the unknown source position. This 
leads to seven free parameters and only five constraints, so from the lens 
modeling itself it is not possible to determine the mass parameters 
without including additional information. Our goal
is to demonstrate that to fit a lensing model introduces new and different
constraints on the mass parameters then derived from rotation curves alone.
For this purposes we will assume some of the mass parameters have been 
determined by some other means (observational or theoretical), and test them
against B1600+434.  Thus we will assume that these
parameters are not being determined from the lensing analysis and thus
do not contribute to the degrees of freedom. This reduces the number
of free parameters to five, and we will show plots where an additional
one or two parameters have been set to a specific value in order to
use $\Delta \chi^2$ as a confidence level estimator.
We minimize the free parameters using the 
Numerical Recipes routine amoeba.f \markcite{press:92}({Press} {et~al.} 1992).  Because our model is
under-constrained it is possible to find combinations of parameters 
that come arbitrarily close to the observables ($\chi^2=0$). These values,
however do not have the statistical significance of a best fit and thus we 
avoid quoting them in this paper.

For the observables we use the flux ratio as measured in the radio, which 
ranges from 1.2 to 1.3.  We adopt the value of $1.25 \pm 0.1$ as 
\markcite{kdj:98}{Koopmans} {et~al.} (1998) did, where the uncertainty includes the 
effect of time variability.
From the NICMOS image we get accurate distances between 
the lens center and the two images; Table \ref{tablepos} lists the positions 
of the quasar images and the companion galaxy (SE) relative to the lens center.
The $x$ axis is taken to lie along the major axis of the disk. Our measured 
separation between the two images $ 1\farcs384 \pm 0\farcs0016$ 
is consistent with that measured in the radio, 
$1\farcs39 \pm 0\farcs01$ \markcite{jack:95}({Jackson} {et~al.} 1995). 
As mentioned above, the lensing model strongly constrains only the 
amplitude and ellipticity of the mass distribution,
so we expect a solution space 
where one can trade off the ellipticity of the halo with more mass in the disk
or bulge.  For clarity we first model 
the system with one PIEMD as is usually done for elliptical lenses.  
This situation can be thought of as the combinations 
of disk, bulge and halo masses that give a perfectly flat rotation curve.  

We treat the companion galaxy 
as a singular isothermal sphere with a circular velocity 200 km s$^{-1}$ as
suggested by its $B$ band luminosity \markcite{kdj:98}({Koopmans} {et~al.} 1998) and at the
redshift of the lens. For comparison we 
include some results that do not include the companion galaxy. 
The effect of the companion galaxy is to increase the combined ellipticity of
the lens galaxy needed to produce the observed image configuration.
\begin{table}
\caption{Distances of QSO images}
\label{tablepos}
\begin{tabular}{lccc}
\tableline
Object & x (arcsec) & y (arcsec) & length (arcsec)\\
\tableline
\tableline
A	&  $0.327\pm0.002$  &  $-0.986\pm0.004$  &  	$1.038 \pm0.0023$\\
B	& $-0.257\pm0.002$  &  $0.269\pm0.002$   &	$0.372 \pm0.0015$\\
SE	&  $0.12\pm0.006$   &  $4.12\pm0.006$	 &  	$4.35  \pm0.004$\\
\tableline
\end{tabular}
\end{table}

\subsubsection{Cosmology}
We do our calculations using $\Omega=1, \Lambda=0$,
and the full beam approximation \markcite{sef}({Schneider}, {Ehlers}, \& {Falco} 1992).
Changing to a different cosmological model has a
small effect on our analysis. The corrections introduced by going to
$\Omega=0.2, \Lambda=0.8$ are less then 15\%.
The largest effect of 
cosmology is determining the angular diameter distance to 
the lens and hence the conversion from arc seconds to kpc.  
At the redshift of the lens 
$1\arcsec=5.5$ kpc for $\Omega=1$ and 6.0 kpc for $\Omega=0.1$ with a Hubble 
constant of 60 km s$^{-1}$ Mpc$^{-1}$. 
Thus, if the cosmological constant is small, all masses and lengths can be 
roughly adjusted to the cosmology of one's choice
by multiplying by $f/5.5$ where $f$ is the number of kiloparsecs that one
arc second corresponds to at the redshift of the lens in the chosen
cosmology. 

\subsection{The one PIEMD fit}  
\begin{figure}[t] % figpie
\vskip .5pc
\begin{minipage}[b]{.46\linewidth}
\centering
\centerline{\epsfig{file=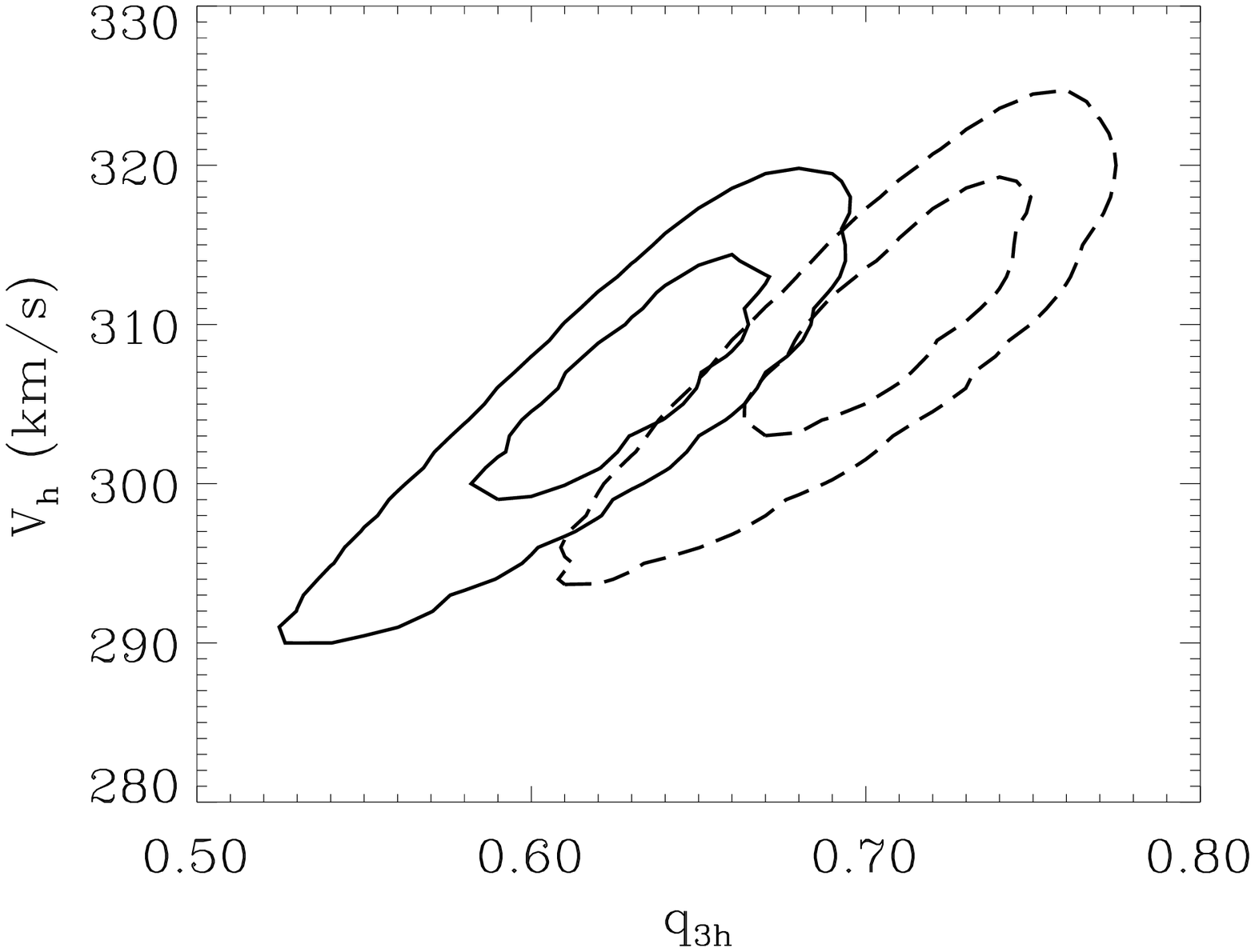,width=\linewidth}}
\end{minipage}\hfill
\begin{minipage}[b]{.46\linewidth}
\centering
\centerline{\epsfig{file=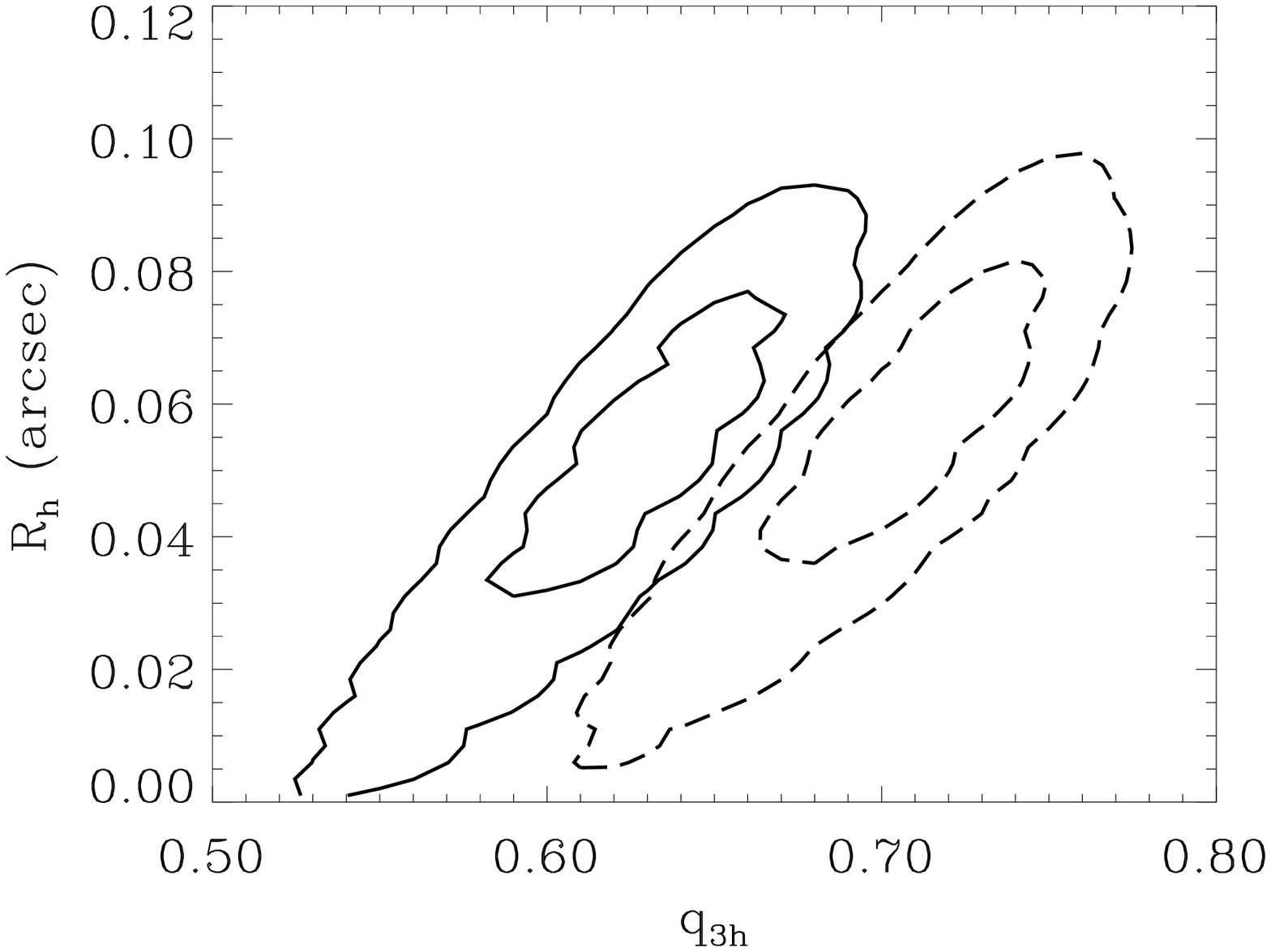,width=\linewidth}}
\end{minipage}\hfill
\vspace{10pt}
\caption{
Shown are the 66\% and 95\% confidence level 
contours into the $V_h$ versus $q_{3h}$ (left panel) and $V_h$ versus $R_h$
(right panel) planes.  The solid lines include the companion galaxy, while 
the dashed lines do not. 
}\label{figpie}
\end{figure}

The PIEMD by itself produces an excellent fit to the image positions and 
flux ratio as has been shown for many elliptical lenses \markcite{kkf:98}({Keeton} {et~al.} 1998). 
Aside from the combined effect of the three mass
components, this model can also be thought of as a mass distribution 
in which the disk and bulge make a negligible contribution.
With the latter perspective, 
this model represents the greatest possible contribution of the halo 
to the total mass, and the maximum ellipticity of the halo, since
the halo solely causes all the lensing.
Contours of $\Delta \chi^2$ of 2.3 and 6.17 corresponding to 68\% and 95\% 
confidence limits for 2 d.o.f. are shown in Figure \ref{figpie}. 
This gives strict upper limits on the halo velocity of 320 km s$^{-1}$ and on
the core radius of $0\farcs1$ or $R_h=0.55 h^{-1}_{60}$ kpc, 
and a lower limit of the dark halo flattening of $q_{3h}= 0.53$, 
in agreement with \markcite{kdj:98}{Koopmans} {et~al.} (1998). Note that this small value for $R_h$ 
implies that the rotation curve of this galaxy does not fall off towards its 
center.  

Also shown are the same contours when the 
companion galaxy is not included in the lensing analysis. 
Since the direction from the lens center to the companion galaxy 
is nearly perpendicular to the plane of the disk, its
contribution to the ellipticity of the projected mass tends to counteract that
of the lens galaxy. Thus the companion galaxy causes the lens galaxy to be
more flattened in order to produce the same image configuration.
The companion galaxy also contributes to the total mass, 
reducing the amount that is associated with the lens galaxy by $\sim 6$ 
km s$^{-1}$.  
It is interesting to note how the rotation velocity and core radius of this 
model are typical for elliptical lenses \markcite{koch:98}(Kochanek {et~al.} 1998), 
confirming our earlier
assertion that an edge-on spiral galaxy lenses like a more massive 
elliptical galaxy.

\subsection{Adding the bulge and disk} 

\begin{figure}[tp] % figdb
\vskip .5pc
\centering
\centerline{\epsfig{file=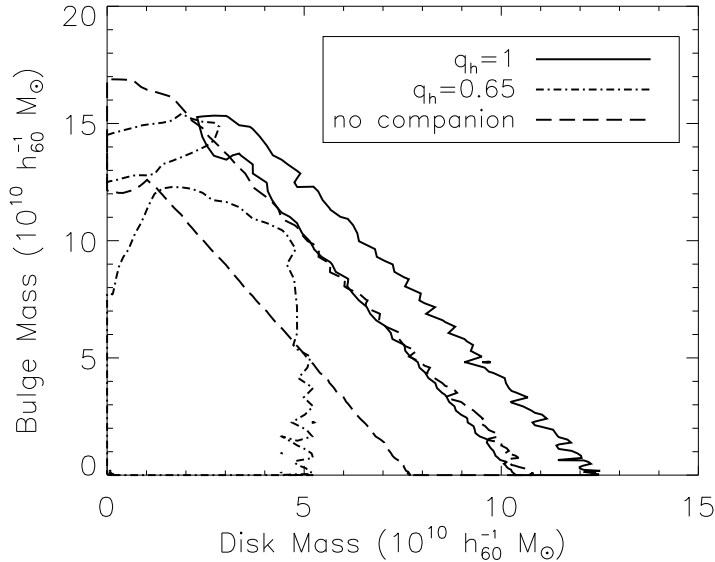,height=.4\linewidth}}
\vspace{10pt}
\caption{95\% confidence contours in the bulge mass versus disk mass plane
for a spherical halo (solid) and  $q_h=0.65$ halo (dash-dot) with the 
companion galaxy, and also
for a spherical halo with no companion galaxy (dash).
}\label{figdb}
\end{figure}

We now add the bulge and disk to our analysis by adding two
chameleon profiles tuned to match the observed exponential profiles of the
disk and bulge.  Because
the scale lengths and axis ratios are measured this introduces only
two new free
parameters: the disk and bulge masses.  We continue to use a PIEMD to represent
the dark halo, so that now there are five mass parameters in the model.  
We thus choose 
to examine two fixed halo ellipticities that bracket the range of solutions and
will leave us with 1 d.o.f.

Figure \ref{figdb} shows the 95\% 
confidence contours ($\Delta \chi^2=4.0$) in bulge mass versus disk
mass for a spherical halo ($q_{3h}=1$), 
a flattened halo ($q_{3h}=0.65$), and also for the
spherical case with no companion galaxy. The effect of the companion galaxy is
to reduce the necessary total projected ellipticity of the lens galaxy, as 
also demonstrated above.  Thus from here on we will conduct our analysis only 
with the companion galaxy included, noting that if the 
companion's redshift or mass is different than we have assumed the effect can 
be compensated by an adjustment in the ellipticity of the dark halo.

Figure \ref{figdb} shows that, as mentioned earlier, there is a large 
degeneracy in the solution parameters.  If we restrict ourselves to
oblate spheroids, the disk and bulge masses are only required to lie below 
$1.5 \times 10^{11} h^{-1}_{60} M_{\sun}$ and 
$1.3 \times 10^{11} h^{-1}_{60} M_{\sun}$ respectively. Although the disk and
bulge masses are not uniquely determined, 
it is clear how they relate to the halo's properties.  
For the spherical halo, $q_{3h}=1$, the ellipticity comes from just
the disk and bulge.  Thus 
a linear relationship exists such that a decrease in 
disk mass must be met with an increase in bulge mass to produce the observed
total ellipticity.  On the other hand, for a flattened halo, $q_{3h}=0.65$, 
the disk is no longer required to match the ellipticity and is limited to 
less then $5 \times 10^{10} h^{-1}_{60} M_{\sun}$.  Instead the bulge and halo
have a linear relationship such that their combined mass 
produces the observed image separation. Thus the upper limit on the bulge 
mass arises when the bulge itself is massive enough to produce the image
separation.

One can ask if there are regions in the solution space where the dark matter 
contribution in the center of the galaxy is small.  It turns out that if the
bulge mass is greater than $1.3 \times 10^{11} h^{-1}_{60} M_{\sun}$,
the core radius of the dark halo may be $\geq 1\farcs3$, but for values of 
the bulge mass less then this, it must be close to zero:
$R_h \leq 0\farcs2$.  The image separation demands a great deal of 
mass in the center of the galaxy, as was shown for the single PIEMD. 
The disk can 
not provide this mass because it increases the combined ellipticity, so if the 
bulge mass is not sufficient, the dark matter must provide the rest.  
Core radii between $0\farcs2-1\farcs3$ are excluded because they produce a flux
ratio inconsistent with the observations. Thus the solutions are easily divided
into two types: those with small (less than $0\farcs2$) and large (greater than
$1\farcs3$) core radii. 

\begin{figure}[t] % figvel
\vskip .5pc
\begin{minipage}[b]{.46\linewidth}
\centering
\centerline{\epsfig{file=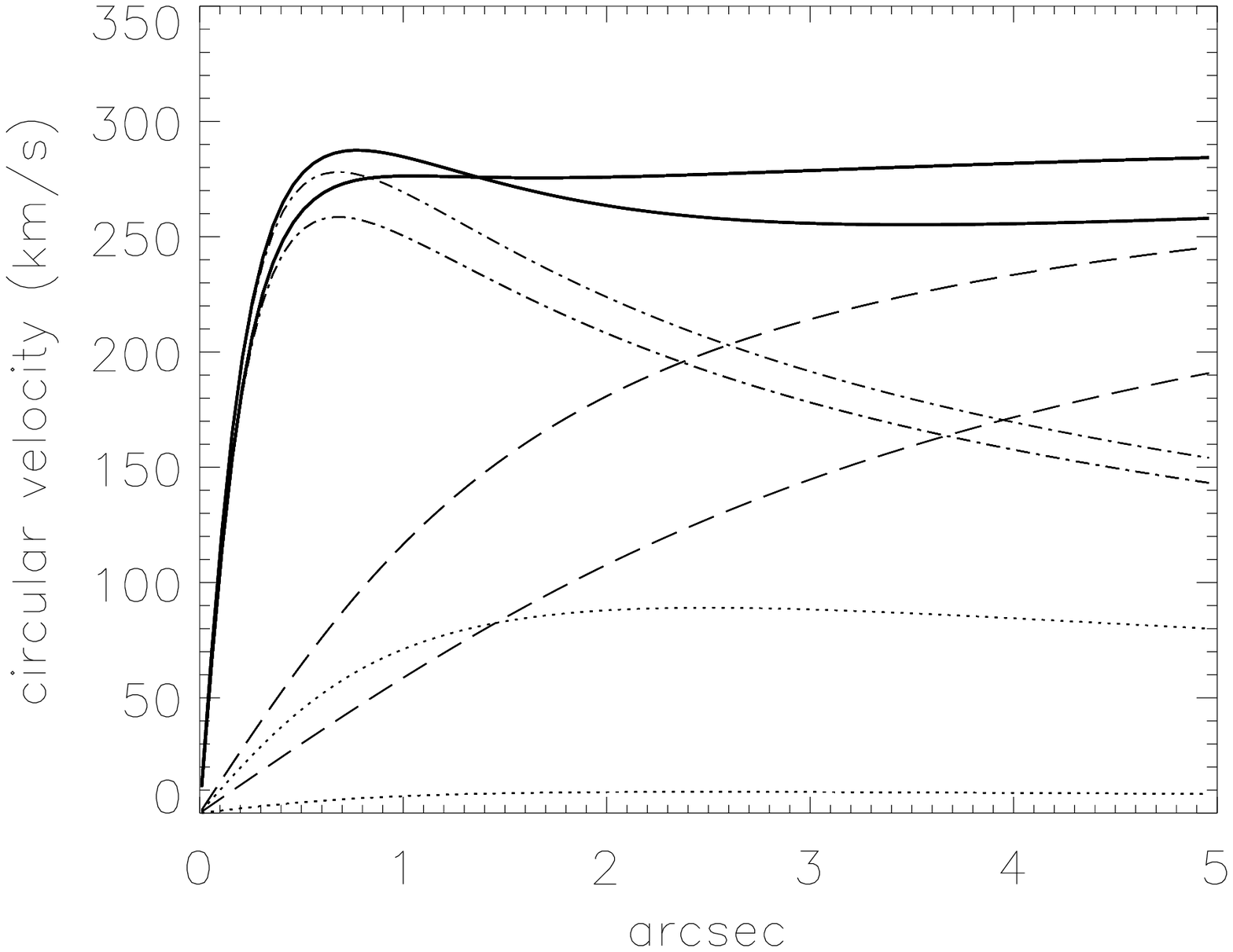,width=\linewidth}}
\end{minipage}\hfill
\begin{minipage}[b]{.46\linewidth}
\centering
\centerline{\epsfig{file=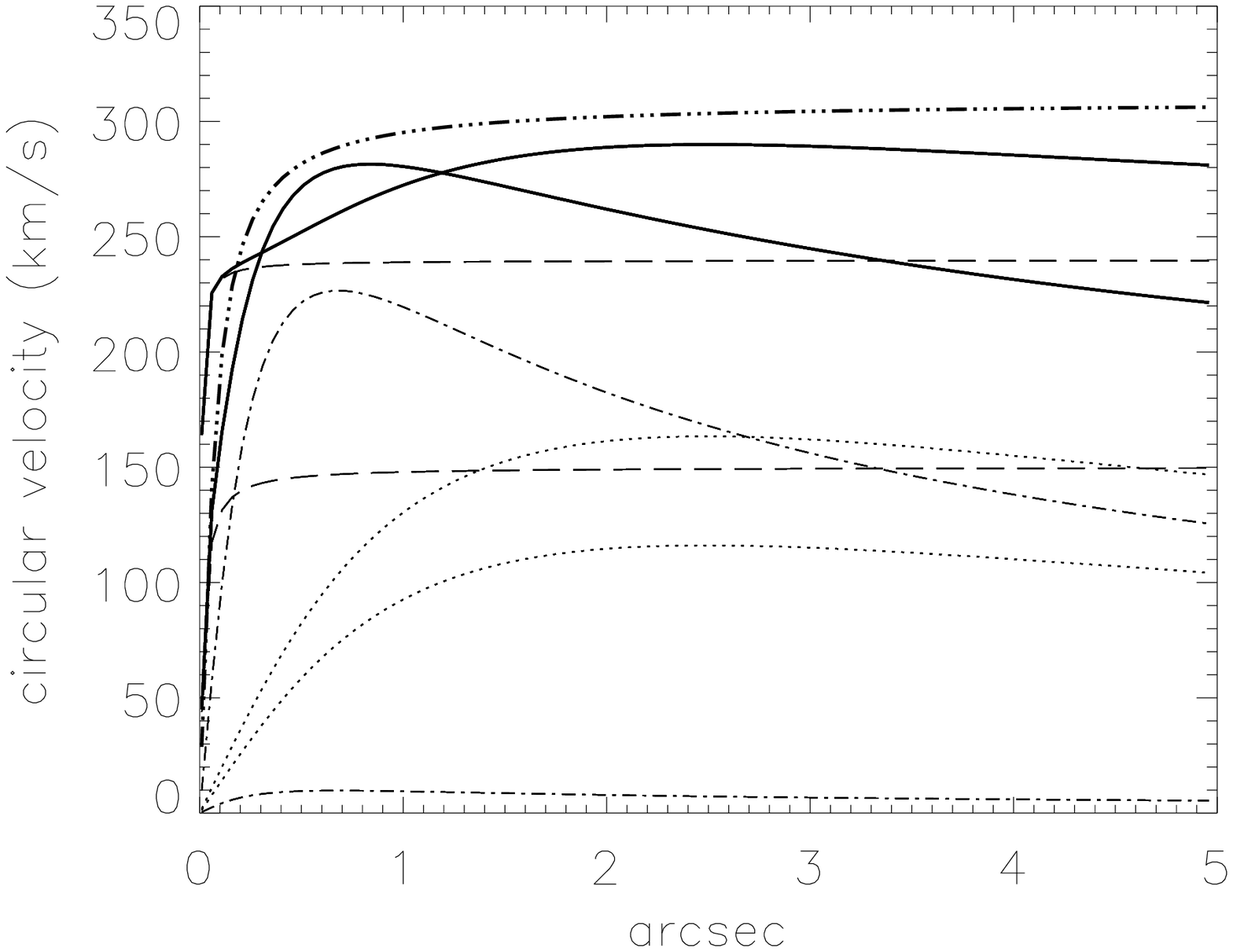,width=\linewidth}}
\end{minipage}\hfill
\vspace{10pt}
\caption{
The rotation curves are plotted for the large (left) and small (right) core
radii solutions from Figure 6.  The lower and upper limits
of disk (dotted), bulge (dash-dot), dark halo (dashed) and combined
(solid) contribution to the rotation curve are shown. 
The right panel shows only the spherical halo, we also include the velocity
profile for the solution of a flattened halo with no disk or bulge contribution
for comparison (dash-dot-dot-dot).
The lensing analysis only constrains the mass inside $1\arcsec$, but we have
assumed the mass models are also valid outside this range, and rejected 
solutions where the rotation curve declines significantly in the outer parts.
}\label{figvel}
\end{figure}

Figure \ref{figvel} shows rotation curves for large, 
$R_h = 7-16 h^{-1}_{60}$ kpc, and small, $R_h \leq 0.5 h^{-1}_{60}$ kpc, 
core radii. 
Since
the more distant image is only $1\arcsec$ from the lens center, we are not
probing out past this distance in our lensing analysis; however,
to compare with
local rotation curves we continue the mass model out to $5\arcsec$.  We 
exclude solutions that do not resemble local rotation curves by showing no 
evidence for dark matter in their outer parts.  This amounts to values of 
$R_h$ greater than $3\arcsec$, or $V_h$ less than 150 kms$^{-1}$.
Each component is represented by two lines in Figure \ref{figvel} 
that bracket the range of values it may have subject to the above restrictions 
and matching the lensing constraints.
In the right panel the halo is kept spherical to show the lower bound on the
disk mass.  If the halo is allowed to be flattened, there is no longer a need for
any disk mass, a solution with just the flattened halo is possible 
(the dash-dot-dot-dot line).
In all cases the rotation curve reaches 270-295 km s$^{-1}$ at $1\arcsec$,
but can vary from 250 to 305 km s$^{-1}$ at $2\arcsec$ depending on which 
component is dominant. These values are well bracketed by the Tully-Fisher 
estimates we derived in Section \ref{b16} and seem to correspond to a
more reasonable amount of extinction correction.  

\subsection{Mass-to-Light Ratios}

\begin{figure}[tp] % figml
\vskip .5pc
\centering
\centerline{\epsfig{file=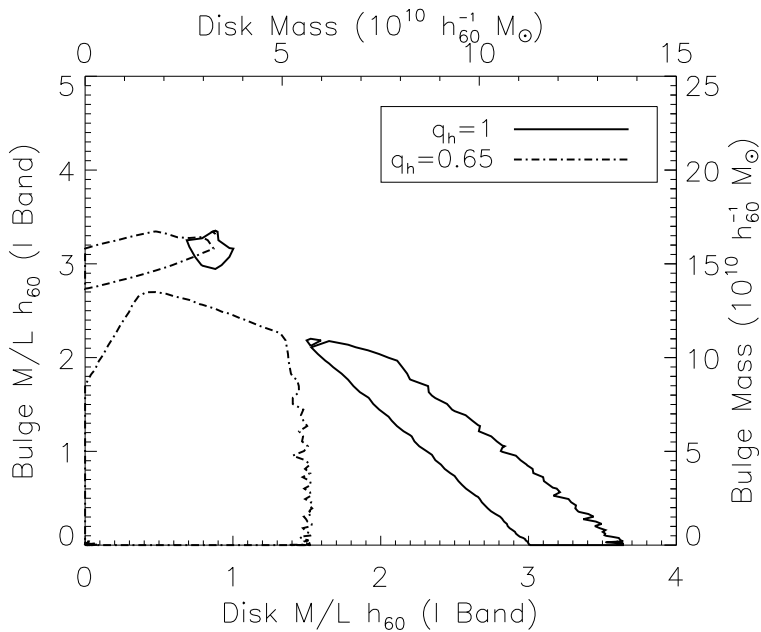,height=.4\linewidth}}
\vspace{10pt}
\caption{Shown are 95\% confidence contours in the mass-to-light ratios 
in solar units of the
bulge versus disk for a spherical halo (solid) and flattened $q_h=0.65$ halo
(dot-dash).
Only solutions that create flat rotation curves are included ($V_h \geq 150$ km
s$^{-1}$ and $R_h \leq 3\arcsec$). No extinction correction is included, so 
the M/L for the disk should be considered an upper limit.
}\label{figml}
\end{figure}

The last piece of information that can be provided by analyzing B1600+434 
is the mass-to-light ratios (M/L) of the bulge and the disk. 
These mass-to-light ratios should reflect the baryonic component in the galaxy
since only baryons cool to form disks and bulges 
in the cold dark matter paradigm.
Figure \ref{figml} shows the mass-to-light ratios of bulge versus disk in 
solar units for the
95\%  
confidence level solutions to the lensing analysis that also produce good 
rotation curves.  The separation between large and small core radii is now
clearly seen, with the large core radius case corresponding to a higher bulge
M/L.  

We have not made any correction to the luminosity for extinction aside from
removing the dust lane, so the values of M/L for the disk should be taken as 
upper limits.  With some knowledge of the mass-to-light ratios of the disk or 
bulge it is possible to further constrain the solutions.  Older stellar 
populations have higher mass-to-light ratios, so we would expect the bulge to 
have a higher mass-to-light ratio than the disk. Extinction in the disk is
probably less then a factor of two, which would exclude solutions with a disk
M/L more then twice the bulge M/L, or disk masses above 
$9 \times 10^{10} h^{-1}_{60} M_{\sun}$.
\markcite{worthey:94}{Worthey} (1994) modeled the mass-to-light ratio for old
and intermediate age stellar populations with a range of metalicities
$-2 \leq$ [Fe/H]$\leq 0.5$. 
His lowest M/L in the $I$ band is 0.9 which provides
a conservative lower limit on the bulge M/L. From Figure \ref{figml} one sees
that this constrains the disk mass to less than 
$10 \times 10^{10} h^{-1}_{60} M_{\sun}$.
\markcite{bott:99}Bottema (1999) derives a disk M/L$=1.6 \pm 0.7$ from stellar velocity 
dispersions of local spiral galaxies.  This seems to disfavor the 
large core radii solutions which have
disk M/L $\leq 1.0$, which is an upper limit as there is 
probably some extinction due to dust.
Thus for this lens system, while considerations 
of the mass-to-light ratio constrain the solution space, 
they do not distinguish between the different types of solution: 
large or small core radius and flat or spherical halo.  

\section{Discussion}

Our analysis of B1600+434 assumes that the mass in the system can be accounted
for by a constant M/L for the disk and the bulge, 
a dark halo that is centered and
has the same orientation as the disk, and mass associated with the companion
galaxy.  If the mass in the bulge or disk does not follow the observed 
light distribution, then we are unable to constrain the model.  Since B1600+434
has a close companion, it is possible that their dark halos have merged to 
form a common halo, centered somewhere besides the center of B1600+434.  
However, since neither galaxy appears disturbed, we feel justified in assuming
that the dark halos are not interacting and are 
thus centered and aligned with the light distribution. 

Our purpose here has been to discuss the importance of spiral lenses in 
determining the relative masses of the disk, bulge, and halo.  We have 
analyzed B1600+434 to give the reader an understanding of how the method works
and also some of its limitations.  Because only the total mass and combined
ellipticity are strongly constrained there will always be degeneracies in the
combination of disk, bulge, and halo masses 
that produce the observed image positions.
Considerations of the flatness of the rotation curve or of the mass-to-light 
ratios can be used to further constrain the solutions.  

Two types of solution seem tenable for the galaxy B1600+434 based on the
lensing analysis, and including our restrictions based on flat rotation curves.
One is a large dark 
halo core radius with a corresponding bulge mass of 
$1.2 - 1.4 \times 10^{11} h^{-1}_{60} M_{\sun}$ and disk mass of $\sim 5 
\times 10^{10} h^{-1}_{60} M_{\sun}$.  The other is a nearly singular 
dark halo with the combined ellipticity coming from either the disk or a 
flattened dark halo. There are three observations that will help to 
restrict the solutions further.  First,
a redshift for the companion galaxy will
establish its contribution to the lensing.  Second, a measure of the 
rotation velocity of B1600+434 can be used to distinguish between 
the different types of solutions.  These solutions produce a circular
velocity at two disk scale lengths ranging from 240 km s$^{-1}$ 
to 310 km s$^{-1}$.  Solutions with a large bulge mass produce lower circular
velocities because the potential falls off the fastest for the bulge. 
The highest values for the circular velocity are reached with the heaviest 
dark halo since its contribution does not fall off. 
Lastly, measuring a time delay for this 
system will reduce uncertainties in the flux ratio of the images,
and measure the potential of the mass distribution.  The models shown in 
Figure \ref{figml} produce time delays ranging from $36-46 h^{-1}_{60}$ days.
The large core radii solutions favor longer time delay $40-46 h^{-1}_{60}$ 
days, but uncertainty in the value of the Hubble constant will make it unlikely
that this measurement will be able to differentiate between models.

Our analysis of B1600+434 cannot determine whether baryons or dark matter
play the more important role in the center of this galaxy.  However, it does 
rule out solutions which have both a large core radius for the dark halo and a 
massive disk. 
Of course the details of 
our results depend upon the radial profile we have assumed for the 
dark halo. 
This, of course, 
is only one galaxy and it would be premature to 
draw any strong conclusions from its analysis. Our point here has not been to 
resolve the disk/halo debate, 
but to demonstrate that spiral galaxy lensing will
contribute valuable information to it.  When a larger sample has been analyzed,
we will learn about mass-to-light ratios of disks and bulges, the flatness 
of the dark halo inside one disk scale length, and the relationship between 
the baryonic and dark components in spiral galaxies.

Acknowledgments: We would like to 
thank Nicole Vogt, whose introduction of AHM and LS made this project possible;
Tsafrir Kolatt for many helpful suggestions during the course of the project 
and James Bullock and Risa Wechsler for stimulating conversions. Also we would 
like to thank Paul Schechter for his kind advice, and the referee, 
Rennan Barkana, whose comments improved the clarity and presentation
of this paper.
AHM acknowledges support from GAANN and a NASA ATP grant at UCSC.
LS gratefully acknowledges financial support from a postdoctoral
fellowship from the Natural Sciences and Engineering Research Council of
Canada.
JH acknowledges support from the Danish Natural Science Research Council (SNF).
AOJ acknowledges support from the Centre for Advanced Study, Oslo.
JRP acknowledges support from NSF and NASA grants at UCSC, and 
RAF acknowledges support from NSF and UM Research Board support at UMSL.
The bibliography was prepared using Jonathan Baker's Astronat package.

\bibliography{}

%% --------------------------------------------------------------------
%% Mon Nov  1 16:27:36 1999
%%   This file was generated automagically from the files
%%   b16.bbl and b16.tex using
%%     ./nat2jour.pl
%%   This file should accompany b16-apj.tex.
%% --------------------------------------------------------------------

\begin{thebibliography}{}

\bibitem[{Andredakis} 1998]{andredakis:98}
{Andredakis}, Y.~C. 1998, \mnras, 295, 725

\bibitem[{Bartelmann} \& {Loeb} 1998]{bl:98}
{Bartelmann}, M. \& {Loeb}, A. 1998, \apj, 503, 48

\bibitem[{Bernstein}, {Guhathakurta},  {Raychaudhury}, {Giovanelli}, {Haynes}, {Herter}, \& {Vogt} 1994]{bern:94}
{Bernstein}, G.~M., {Guhathakurta}, P., {Raychaudhury}, S., {Giovanelli}, R.,  {Haynes}, M.~P., {Herter}, T., \& {Vogt}, N.~P. 1994, \aj, 107, 1962

\bibitem[{Blain}, {M\"{o}ller}, \& {Maller} 1999]{bmm:99}
{Blain}, A.~W., {M\"{o}ller}, O., \& {Maller}, A.~H. 1999, \mnras, 303, 423

\bibitem[Bosma 1978]{bosma:78}
Bosma, A. 1978, PhD thesis, University of Groningen

\bibitem[{Bosma} 1998]{bosma:98}
{Bosma}, A. 1998, in Galaxy Dynamics, To appear in ASP Conference Series, ed.  D.~R. Merritt, M.~Valluri, \& J.~A. Sellwood, E21, astro-ph/9812013

\bibitem[Bottema 1999]{bott:99}
Bottema, R. 1999, astro-ph/9902240

\bibitem[{Brainerd}, {Blandford}, \&  {Smail} 1996]{bbs:96}
{Brainerd}, T.~G., {Blandford}, R.~D., \& {Smail}, I. 1996, \apj, 466, 623

\bibitem[{Browne} 1998]{browne:98}
{Browne}, I. 1998, in Observational Cosmology with the New Radio Surveys,  Dordrecht: Kluwer Academic Publishers, Astrophysics and space science library  (ASSL) Series vol no: 226, 323

\bibitem[{Courteau}, {De Jong}, \&  {Broeils} 1996]{cdb:96}
{Courteau}, S., {De Jong}, R.~S., \& {Broeils}, A.~H. 1996, \apjl, 457, L73

\bibitem[{Courteau} \& {Rix} 1999]{cr:99}
{Courteau}, S. \& {Rix}, H.~W. 1999, \apj, 513, 561

\bibitem[{Dalcanton}, {Spergel}, \&  {Summers} 1997]{dss:97}
{Dalcanton}, J.~J., {Spergel}, D.~N., \& {Summers}, F.~J. 1997, \apj, 482, 659

\bibitem[{De Jong} 1994]{dejong:94}
{De Jong}, R.~S. 1994, PhD thesis, University of Groningen

\bibitem[de~Vaucoulers 1948]{deV:48}
de~Vaucoulers, G. 1948, Ann. Astrophys., 247, 11

\bibitem[{Dell'Antonio} \& {Tyson} 1996]{dt:96}
{Dell'Antonio}, I.~P. \& {Tyson}, J.~A. 1996, \apjl, 473, L17

\bibitem[{Dubinski} \& {Carlberg} 1991]{dc:91}
{Dubinski}, J. \& {Carlberg}, R.~G. 1991, \apj, 378, 496

\bibitem[{Fassnacht} \& {Cohen} 1998]{fc:98}
{Fassnacht}, C.~D. \& {Cohen}, J.~G. 1998, \aj, 115, 377

\bibitem[{Fruchter} \& {Hook} 1998]{fh:98}
{Fruchter}, A.~S. \& {Hook}, R.~N. 1998, astro-ph/9808087

\bibitem[Gonzales, Williams, Bullock, Kolatt, \&  Primack 1999]{gonzales:99}
Gonzales, A.~H., Williams, K.~A., Bullock, J.~S., Kolatt, T.~S., \& Primack,  J.~R. 1999, \apj, submitted

\bibitem[Hjorth \& Kneib 1999]{hk:99}
Hjorth, J. \& Kneib, J.~P. 1999, \apj, submitted

\bibitem[{Hudson}, {Gwyn}, {Dahle}, \&  {Kaiser} 1998]{hudson:98}
{Hudson}, M.~J., {Gwyn}, S. D.~J., {Dahle}, H., \& {Kaiser}, N. 1998, \apj,  503, 531

\bibitem[{Jackson}, {De Bruyn}, {Myers}, {Bremer},  {Miley}, {Schilizzi}, {Browne}, {Nair}, {Wilkinson}, {Blandford}, {Pearson},  \& {Readhead} 1995]{jack:95}
{Jackson}, N., {De Bruyn}, A.~G., {Myers}, S., {Bremer}, M.~N., {Miley}, G.~K.,  {Schilizzi}, R.~T., {Browne}, I. W.~A., {Nair}, S., {et al.} 1995, \mnras,  274, L25

\bibitem[{Jaunsen} \& {Hjorth} 1997]{jh:97}
{Jaunsen}, A.~O. \& {Hjorth}, J. 1997, \aap, 317, L39

\bibitem[{Kassiola} \& {Kovner} 1993]{kk:93}
{Kassiola}, A. \& {Kovner}, I. 1993, \apj, 417, 450

\bibitem[{Keeton} \& {Kochanek} 1998]{kk:98}
{Keeton}, C.~R. \& {Kochanek}, C.~S. 1998, \apj, 495, 157

\bibitem[{Keeton}, {Kochanek}, \& {Falco} 1998]{kkf:98}
{Keeton}, C.~R., {Kochanek}, C.~S., \& {Falco}, E.~E. 1998, \apj, 509, 561

\bibitem[Kochanek, Falco, Impey, Lehar, McLeod, \&  Rix 1998]{koch:98}
Kochanek, C.~S., Falco, E.~E., Impey, C.~D., Lehar, J., McLeod, B.~A., \& Rix,  H.-W. 1998, astro-ph/9811111

\bibitem[{Koopmans}, {De Bruyn}, \&  {Jackson} 1998]{kdj:98}
{Koopmans}, L. V.~E., {De Bruyn}, A.~G., \& {Jackson}, N. 1998, \mnras, 295,  534

\bibitem[{Kormann}, {Schneider}, \&  {Bartelmann} 1994]{ksb:94}
{Kormann}, R., {Schneider}, P., \& {Bartelmann}, M. 1994, \aap, 284, 285

\bibitem[{Loeb} 1996]{loeb:96}
{Loeb}, A. 1996, in Eighteenth Texas Symposium on Relativistic Astrophysics and  Cosmology, ed. A.~V. Olinto, J.~A. Frieman, \& D.~N. Schramm (Singarpore:  World Scientific), 465

\bibitem[Maller 1999]{maller:99}
Maller, A.~H. 1999, PhD thesis, Univ. California, Santa Cruz

\bibitem[{Maller}, {Flores}, \& {Primack} 1997]{mfp:97}
{Maller}, A.~H., {Flores}, R.~A., \& {Primack}, J.~R. 1997, \apj, 486, 681

\bibitem[{Maller}, {Flores}, \& {Primack} 1998]{mfp:98}
{Maller}, A.~H., {Flores}, R.~A., \& {Primack}, J.~R. 1998, in ASP Conference  Series 136: Galactic Halos, ed. D.~Zaritsky, 311

\bibitem[{Marleau} \& {Simard} 1998]{ms:98}
{Marleau}, F.~R. \& {Simard}, L. 1998, \apj, 507, 585

\bibitem[{Mo}, {Mao}, \& {White} 1998]{mmw:98}
{Mo}, H.~J., {Mao}, S., \& {White}, S. D.~M. 1998, \mnras, 295, 319

\bibitem[{M\"{o}ller} \& {Blain} 1998]{mb:98}
{M\"{o}ller}, O. \& {Blain}, A.~W. 1998, \mnras, 299, 845

\bibitem[Munoz, Falco, Kochanek, Lehar, McLeod, Impey,  Rix, \& Peng 1999]{munoz:99}
Munoz, J.~A., Falco, E.~E., Kochanek, C.~S., Lehar, J., McLeod, B.~A., Impey,  C.~D., Rix, H.-W., \& Peng, C.~Y. 1999, astro-ph/9902131

\bibitem[{Narayan} \& {Bartelmann} 1999]{nb:96}
{Narayan}, R. \& {Bartelmann}, M. 1999, in Formation of Structure in the  Universe, ed. A.~Dekel \& J.~P. Ostriker (Cambridge: University Press),  360--432

\bibitem[{Natarajan}, {Kneib}, {Smail}, \&  {Ellis} 1998]{nata:98}
{Natarajan}, P., {Kneib}, J.~P., {Smail}, I., \& {Ellis}, R.~S. 1998, \apj,  499, 600

\bibitem[{Poggianti} 1997]{pogg:97}
{Poggianti}, B.~M. 1997, \aaps, 122, 399

\bibitem[{Press}, {Teukolsky}, {Vetterling}, \&  {Flannery} 1992]{press:92}
{Press}, W.~H., {Teukolsky}, S.~A., {Vetterling}, W.~T., \& {Flannery}, B.~P.  1992, Numerical recipes in FORTRAN. The art of scientific computing, 2nd edn.  (Cambridge: University Press)

\bibitem[{Rubin}, {Thonnard}, \& {Ford} 1978]{rtf:78}
{Rubin}, V.~C., {Thonnard}, N., \& {Ford}, W.~K., J. 1978, \apjl, 225, L107

\bibitem[{Sackett} 1998]{sack:98}
{Sackett}, P.~D. 1998, in Galaxy Dynamics, To appear in ASP Conference Series,  ed. D.~R. Merritt, M.~Valluri, \& J.~A. Sellwood, E26, astro-ph/9903420

\bibitem[{Sallucci} \& {Persic} 1999]{sp:99}
{Sallucci}, P. \& {Persic}, M. 1999, \aap, in press, astro-ph/9903432

\bibitem[{Schneider}, {Ehlers}, \& {Falco} 1992]{sef}
{Schneider}, P., {Ehlers}, J., \& {Falco}, E.~E. 1992, Gravitational Lenses  (Berlin: Springer-Verlag)

\bibitem[{Sellwood} 1998]{sell:98}
{Sellwood}, J. 1998, in Galaxy Dynamics, To appear in ASP Conference Series,  ed. D.~R. Merritt, M.~Valluri, \& J.~A. Sellwood, E22, astro-ph/9903184

\bibitem[{Simard} 1998]{simard:98}
{Simard}, L. 1998, ASP Conference Series 145: Astronomical Data Analysis  Software and Systems VII, 7, 108

\bibitem[Somerville \& Primack 1999]{sp:98}
Somerville, R.~S. \& Primack, J.~R. 1999, \mnras, accepted, astro-ph/9802268

\bibitem[{Tully} \& {Fisher} 1977]{tf:77}
{Tully}, R.~B. \& {Fisher}, J.~R. 1977, \aap, 54, 661

\bibitem[{Turner}, {Ostriker}, \& {Gott} 1984]{tog:84}
{Turner}, E.~L., {Ostriker}, J.~P., \& {Gott}, J.~R., I. 1984, \apj, 284, 1

\bibitem[{Van Albada} \& {Sancisi} 1986]{vs:86}
{Van Albada}, T.~S. \& {Sancisi}, R. 1986, Royal Society of London  Philosophical Transactions Series, 320, 447

\bibitem[{Wang} \& {Turner} 1997]{wt:97}
{Wang}, Y. \& {Turner}, E.~L. 1997, \mnras, 292, 863

\bibitem[{Warren}, {Quinn}, {Salmon}, \&  {Zurek} 1992]{warr:92}
{Warren}, M.~S., {Quinn}, P.~J., {Salmon}, J.~K., \& {Zurek}, W.~H. 1992, \apj,  399, 405

\bibitem[{Worthey} 1994]{worthey:94}
{Worthey}, G. 1994, \apjs, 95, 107

\end{thebibliography}

\end{document}